\documentclass{aa}
\bibpunct{(}{)}{;}{a}{}{,} 
\usepackage{graphicx}
\usepackage{txfonts}
\usepackage{color}

\begin{document}

\title{What brakes the Crab pulsar?}

 \author{A.~\v{C}ade\v{z}
          \inst{1}
          \and
          L.~Zampieri\inst{2}
          \and
          C.~Barbieri\inst{2,3}
          \and
          M.~Calvani\inst{2}
          \and
          G.~Naletto\inst{2,4,5}
          \and
          M.~Barbieri\inst{2,3,6}
          \and
          D.~Ponikvar\inst{1}
          }

   \institute{Faculty of Mathematics and Physics, University of Ljubljana,
              Slovenia\\
              \email{Andrej.Cadez@fmf.uni-lj.si}
         \and
             INAF, Astronomical Observatory of Padova, Italy
          \and
             Department of Physics and Astronomy, University of Padova, Italy
             \and
             Department of Information Engineering, University of Padova, Italy
             \and
             CNR-IFN UOS Padova LUXOR, Padova, Italy
             \and
             Department of Physics, University of Atacama, Copiapo, Chile             }

\date{Received ??, 2015; accepted ??, 2015}

\abstract
{Optical observations provide convincing evidence that the optical
phase of the Crab pulsar follows the radio one closely.
Since optical data do not depend on dispersion measure variations,
they provide a  robust and independent confirmation of the radio timing solution.}
{The aim of this paper is to find a global mathematical description of Crab pulsar's phase as a function of time for the complete set of published Jodrell Bank radio ephemerides (JBE) in the period 1988 – 2014.}{We apply the mathematical techniques developed for analyzing optical observations to the analysis of JBE. We break
the whole period into a series of episodes and express the phase of the pulsar in each episode as the sum of two analytical functions. The first function is the best-fitting local braking index law, and the second function represents small residuals from this law with an amplitude of only a few turns, which rapidly relaxes to the local braking index law.}
{From our analysis, we demonstrate  that the power law
index undergoes ``instantaneous'' changes at the time of observed
jumps in rotational frequency (glitches). We find that the phase
evolution of the Crab pulsar is dominated by a series of constant
braking law episodes, with the braking index changing abruptly after
each episode in the range of values between 2.1 and 2.6. Deviations
from such a regular phase description behave as oscillations
triggered by glitches and amount to fewer than 40 turns during the
above period, in which the pulsar has made more than 2x10$^{10}$ turns. }{Our analysis does not favor the explanation that glitches are connected to
phenomena occurring in the interior of the pulsar. On the contrary,
timing irregularities and changes in slow down rate seem to point
to electromagnetic interaction of the pulsar with the surrounding
environment.}

\maketitle

\section{Introduction}

Pulsars are rotating neutron stars with a magnetic field misaligned with
respect to their rotation axis. The periodic pulse of light that we see is
caused by a beam of radiation produced in the magnetosphere and pointing in
our direction at each rotation.
Such a rotating magnetic dipole loses energy at the
expense of its rotational energy so decelerates in time. The
mechanism by which this takes place has been investigated ever since the
discovery of pulsars but has not yet been completely understood.
The idea that
the slowdown mechanism consists of a radiative torque on a rotating
magnetic dipole and of the torque by which the pulsar drives the pulsar
wind was proposed early \citep{b1}, together with the idea that
the braking torque is described by a power-law relation between the
frequency of rotation $f$ and its derivative: $\dot f = - K f^{n}$.

The braking index $n = f \ddot f/{\dot f}^2$ is expected to be three in the case of
braking by pure dipole radiation and one in the case of a pulsar wind dominated
torque \citep{b69}. This proposal has stimulated a long-lasting effort to determine and
classify pulsars by their braking indices. After 23 years of
observations  of the Crab pulsar \citet{b2}
found that, between sudden jumps in
rotational frequency (glitches), the rotational slowdown is described well
by a power law with $n = 2.51\pm 0.01$.
This braking index does not hint
at any simple model for the braking mechanism.
Long-term well-defined values for the braking
index have been confirmed for four other pulsars \citep{b3}, and
they consistently give $ n < 3$, suggesting that magnetic dipole radiation
alone is not sufficient to account for the observed spin-down.

The study of the pulsar braking mechanism is made difficult by known
irregularities in pulsar clocks, which take the form of sudden jumps in
rotational frequency (glitches) or {more gradual} deviations from regular spin-down (timing
noise). Recent systematic analyses of timing noise of the Crab and other
pulsars show that it is not random, as was assumed until quite
recently \citep{b4}. The comparison of a large number of pulsars
has demonstrated that the evolution of the rotational phase of young pulsars is
dominated by long ($\sim 1$ yr) relaxation periods following the
occurrence of significant glitches, whereas older pulsars show
quasi-periodic behavior with phase modulations on typical timescales
between $\sim 1$ and 10 yr \citep{b5}. Such quasi-periodic
changes are sometimes correlated with { discrete} variations in slowdown rate $\dot f$
and pulse profile \citep{b6}. These properties have been interpreted
in terms of phenomena occurring in pulsar magnetospheres \citep{b6}. The
periodically active pulsar PSR B1931$+$24, which exhibits on- and of-switching of the radio emission and drastic changes in braking torque, was
proposed as a prototype of a pulsar-magnetosphere interaction system, in
which a varying flow of charged particles drives the braking
mechanism \citep{b7}. Fluctuations in the size of acceleration gaps were
also considered as possible sources of variations in particle current flow
and braking torque \citep{b8}.

Pulsar glitches have not been considered by most authors as part of the same
mechanism, but as a different phenomenon that is connected with internal pulsar
dynamics \citep{b9,b10,b11,b12}. Indeed, the persistent increase in
slowdown rate of the Crab pulsar from 1969 through 1993 \citep{b2} was
interpreted as being caused by a decrease in the moment of inertia due to
interaction of the internal superfluid core with the crust of the neutronstar. An alternative explanation in terms of torque increase was discarded,
because it was expected to be accompanied by a change in the configuration
of pulsar's magnetic field that would likely induce a change of pulse
profile and of the braking index. This was not found in measurements taken
in 1969-1993 \citep{b2}.

In this paper we analyze the phase history of the Crab pulsar and
find a very accurate mathematical description of its behavior.
Such unifying description indicates that, in our opinion, glitches  and timing noise
are part of the same braking mechanism that undergoes sudden
changes during glitches. {Preliminary results of this analysis were
presented by one of us (A\v C) at the Prague Synergy 2013 Conference (``Interaction of a pulsar
with the surrounding nebula: the case of Crab''\footnote{http://www.synergy2013.physics.cz/index.php/programes}).

Very recently, \cite{b51} have published results of a similar analysis using
45 years of radio data on the rotational history of the Crab pulsar.
We will compare their method with ours, pointing out similarities and differences and,
more important, giving a different interpretation  of the
observed phenomenology}.
The plan of the paper is as follows.
Section~2 touches on questions raised by optical timing observations of the Crab
pulsar and their comparison with the Jodrell Bank radio ephemerides (JBE, http://www.jb.man.ac.uk/$\sim$pulsar/crab.html). Section~3 deals with phase analysis of JBE
data and discusses the changing braking law index. Section~4
contains an analysis of phase residuals and completes the
description of the evolution of rotational phase of the Crab pulsar.
 Conclusions follow in Section~5. Some more technical
details are presented in the Appendix.

\section{Optical timing of the Crab pulsar}

During the period from the end of 2008 through the end of 2009,
which was characterized by the absence of significant glitches in the JBE, we obtained three sets of high signal-to-noise ratio optical observations of the Crab pulsar. The first set of data was obtained in October 2008 with the ultra-fast photon counter Aqueye \citep{b13},
mounted on the Copernico telescope at Asiago, while the second and third sets were taken with a similar instrument, Iqueye
\citep{b14}, mounted on the ESO New Technology Telescope at La Silla in January and December 2009.

We measured optical pulse arrival times during two-second intervals by correlating the incoming photon rate with the average pulse profile and define the starting point for the optical phase at the maximum of the main pulse as defined by the template. Optical phase residuals during a typical observation run are Gaussian-distributed with the width consistent with photon-counting noise.
In this way a typical one-hour observation yields a local phase model with statistical phase uncertainty of $\sim$ 0.3$\mu$s at Le Silla and $\sim$ 0.45$\mu$s at Asiago \citep{b15,b16}.

\begin{figure*}
\centering
\includegraphics[width=18cm]{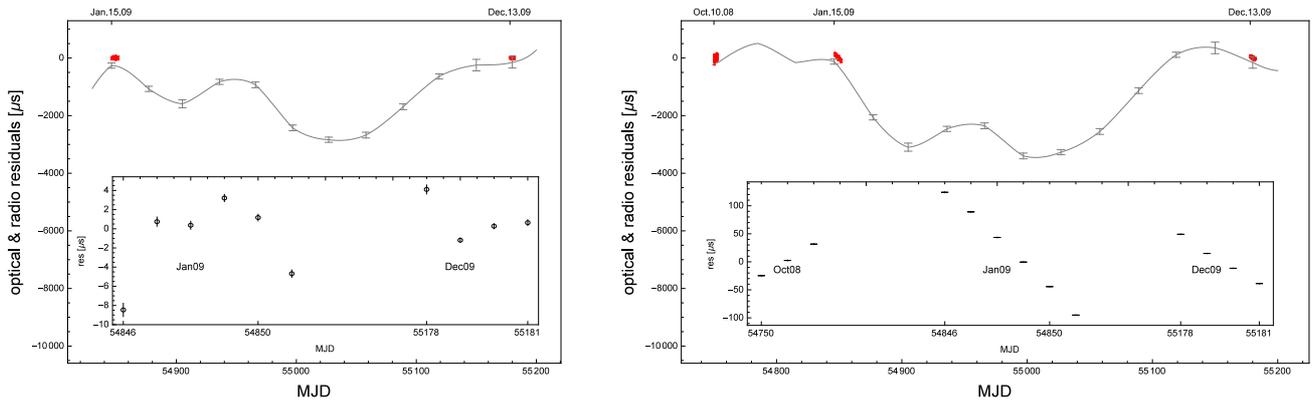}
\caption{Residuals of optical and radio phase with respect to a BIM model.
Left: best-fitting BIM to 2009 Iqueye data ($n = 2.437$).
Right: best-fitting BIM to 2008 - 2009 Aqueye and Iqueye data ($n = 2.476$). Optical
residuals are plotted as red data points  clustered at the
time of optical
observations, while Jodrell Bank residuals (over the same interval of time)
are represented by a gray line crossed by error bars at epoch dates of JBE.
Insets: Zoom of average optical phase residuals during observation nights with 1$\sigma$
error bars. The scales of the y-axis in the insets are very different.  }
\label{Fig1}
\end{figure*}

Since our data were taken with two different instruments at very different locations on Earth, we chose to analyze them independently.
Our analysis of optical data is illustrated in Fig.~\ref{Fig1}, where JBE radio phase residuals (calculated as described in the next section) are also shown.
It turned out that the Iqueye data fit a braking index model (BIM, see below) with an unexpectedly high precision, i.e. the frequency and frequency derivative determined from January and December 2009 data are so tightly constrained that essentially a single braking index law solution with $n=2.437$ also fixes the relative phase between January and December
(left panel in Fig.~\ref{Fig1}).
This leads to a phase description with respect to which the optical data points deviate by less then 10$\mu$s. Thus, the 2009 $n=2.437$ braking law solution appears to be a good baseline for measuring the 2009 Crab phase noise, which is clearly detected at the microsecond level during this quiet period of Crab history.
The phase predicted by this solution agrees (to within the well-documented $150-250 \mu$s radio phase lag) with radio ephemerides phase on the dates  of when we gathered our data and differs from radio ephemerides on the whole interval by less than 3~ms \citep{b16}.
The Asiago October 2008 data were added to the analysis after checking the consistency of our timing protocols on the two sites and the equality of timing response of the two instruments. These data are again consistent with radio ephemerides, but do not follow the 2009 $n=2.437$ braking index law as well, because they miss the prediction by almost 8~ms. It turns out that our complete set of optical data can be fitted by a $n=2.476$ braking index law with respect to which radio phase differs by less than 4~ms in the whole 14-month interval
(right panel in Fig.~\ref{Fig1}). However, in this case the optical phase distinctly shows large excursions (up to 50$~\mu$s per day) with respect to this braking index law. The good fit of our January and December data to the braking index law appears as a rare coincidence, but it stimulated us to ask the question whether there may be a natural baseline with respect to which pulsar phase noise should be measured and how well and for how long a braking index law can approximate the phase history of the Crab pulsar.

\begin{figure*}
\centering
\includegraphics[width=17cm]{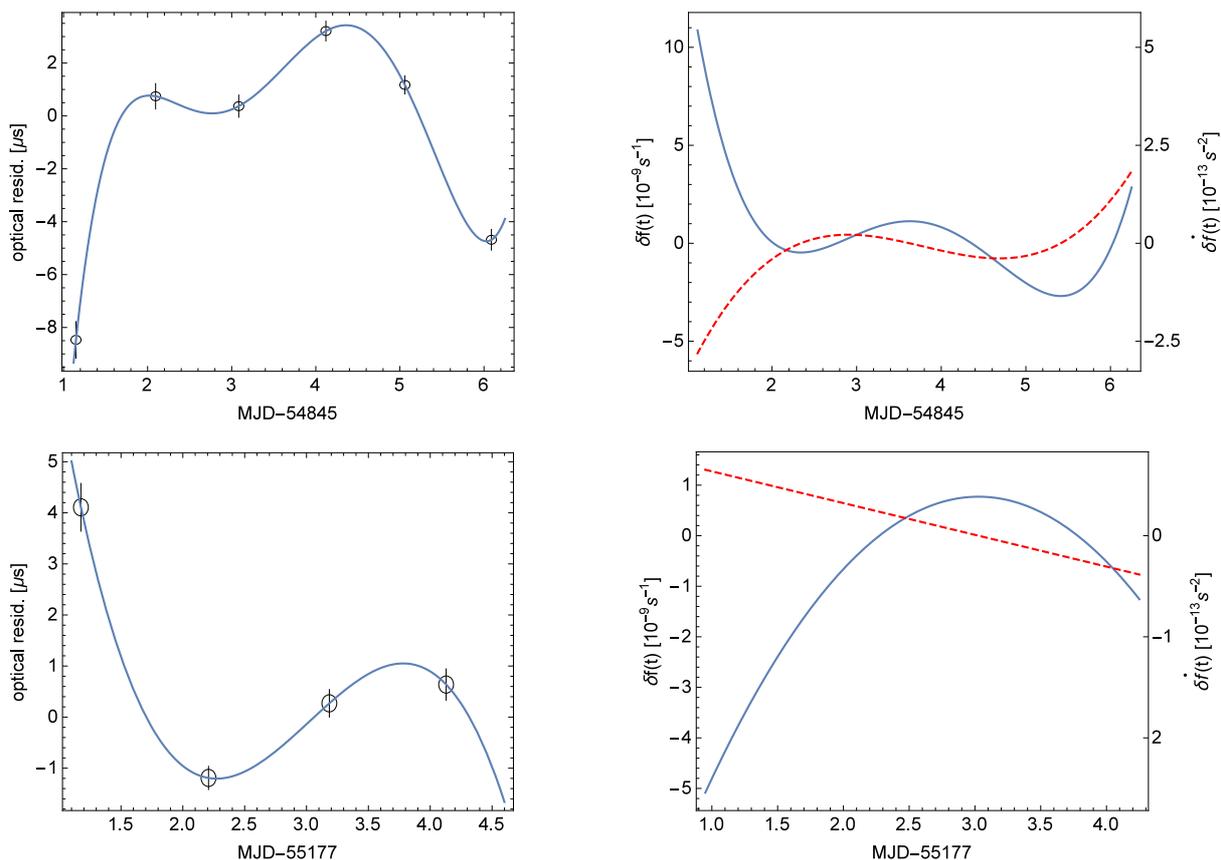}
\caption{January (top) and December (bottom) 2009 optical residuals (Fig.~\ref{Fig1}) fitted to polynomials that reveal ``typical'' noise in pulsar frequency and frequency derivative, as shown on the right, in solid line for frequency residuals (left scale) and dashed for frequency derivative residuals (right scale).}
\label{Fig2}
\end{figure*}

Our analysis leads to the following conclusions:
\begin{itemize}
\item
 The optical phase always leads the radio one. For three sets of observations, the time lapse between optical measurement and JBE reported radio phase  (the latter refers to the center of the main pulse) was between 160 and 260$\mu$s \citep{b15,b16}. Therefore, there is no evidence for any significant change in the delay between the three epochs. Accurate estimates of the X-ray-radio delay have been reported by \citet{rots2004} (344$\pm$40 $\mu$s) and show that it does not change on a timescale of several years. No other difference between optical and radio outside the quoted interval of uncertainty has been found. We note that this delay is also consistent with other recent measurements performed in the optical band (e.g., \citealt{b17,b18}).
\item
A braking index law can reproduce the phase history during the studied 14-month period to within one turn with a narrow range of braking parameters, yet residuals with respect to such a description represent timing noise present on all wavebands and thus most likely reflect genuine variations in rotation of the Crab pulsar. This then motivates us to use JBE data for studying the long-term intrinsic rotation properties of the pulsar.
\item
The small residuals shown in the inset of lefthand part of Fig.~\ref{Fig1} are real and should be considered as typical of short, daily
timescale phase noise during a quiet glitch-less period of pulsar history.
They are shown again in
Fig.~\ref{Fig2}, together with polynomial fits through data points on the left, and frequency and frequency derivative residuals corresponding to these polynomials on the right. Thus, these data suggest ``typical'' frequency noise on a daily scale at the level of $\sim 10^{-8}s^{-1}$ and ``typical'' frequency derivative noise on a daily scale at the level of a few times $10^{-13}s^{-2}$. These estimates are in reasonable agreement  with the difference in frequency and frequency derivative derived from optical and radio data \citep{b16} which are\hfill\break January 15. 2009: \hfill\break $f_{opt}-f_{radio}=-5.64\times 10^{-9}s^{-1}$, $\dot f_{opt}-\dot f_{radio}=1.55\times 10^{-14}s^{-2}$;\hfill \break and December 15. 2009: \hfill\break $f_{opt}-f_{radio}=-6.45\times 10^{-9}s^{-1}$, $\dot f_{opt}-\dot f_{radio}=-6.86\times 10^{-15}s^{-2}$.
\end{itemize}

\subsection{Braking index model implementation}

{ The braking index model implies that the phase $\varphi$ can be expressed as}
\begin{equation}
\varphi(t) = c + a(1+b t)^s,
\label{EQN1}
\end{equation}
{where $c$ is an integration constant that can be used to adjust the initial phase}, $a$ and $b$ are
parameters that are related to the age and frequency of the pulsar
at $t=0$, and $s =\frac{n-2}{n-1}$ is the braking parameter. The
frequency and its derivative are then the following functions of
time $f = \dot\varphi=s a b (1+b t)^{s-1}$ and $\dot
f=\ddot\varphi=s(s-1) a b^2 (1+b t)^{s-2}$. The last equation leads
to the familiar form of the braking index law:
 \begin{equation}
\dot f=\left [b(s-1)\left (s a b \right )^\frac{1}{s-1} \right]f^\frac{s-2}{s-1}
\label{EQN2}
,\end{equation}
therefore $n=\frac{s-2}{s-1}$.

To compare our optical data with radio data as documented in JBE, we devised a numerical procedure that expresses the phase as the number of turns that the pulsar has made since the first pulse arrival time on May 15. 1988; i.e., $\phi(t)=0$ at MJD=47296.0000003712.  More details are given in Appendix A.

\section{Braking episodes}

\begin{figure*}
\centering
\includegraphics[width=17cm]{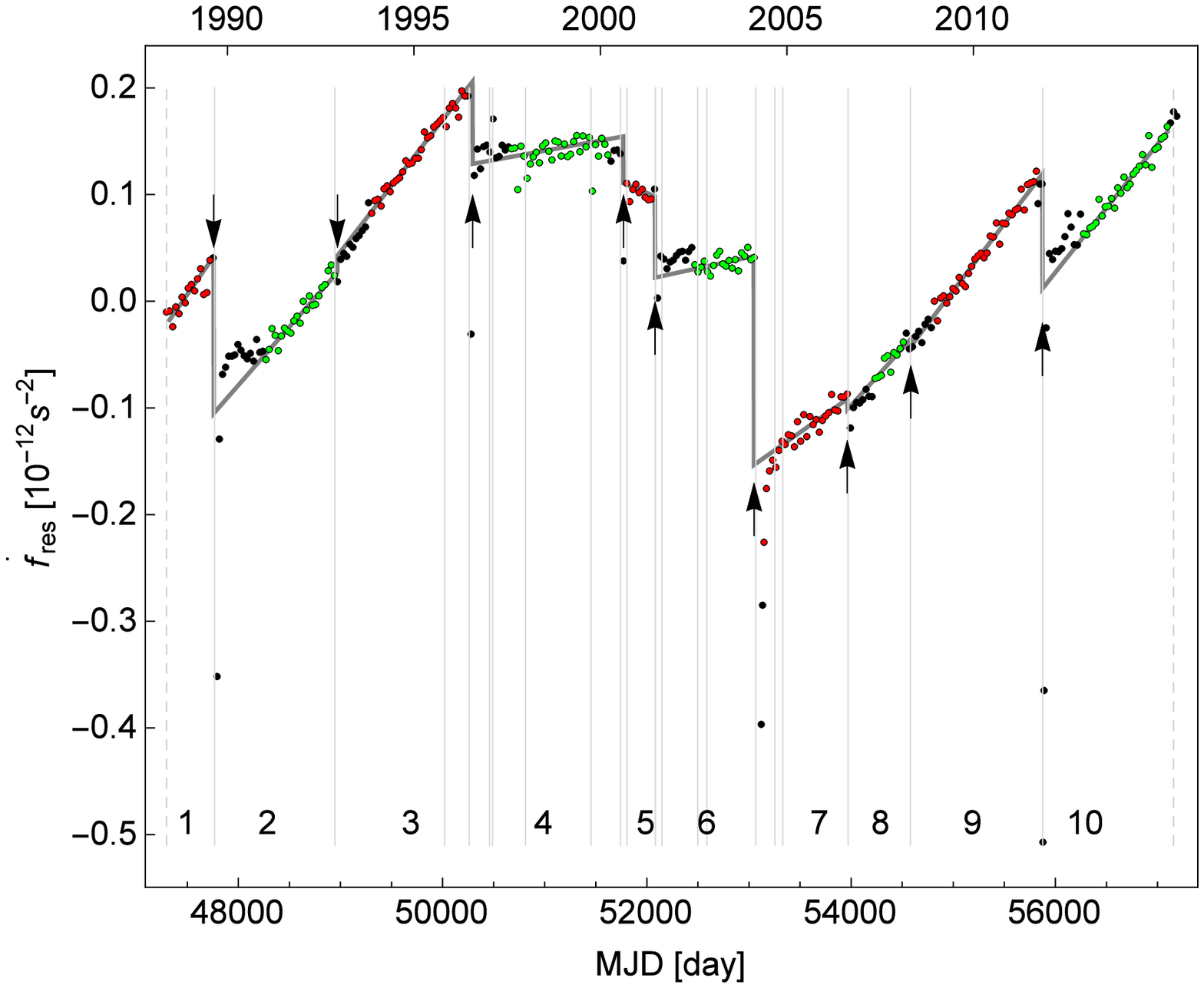}
\caption{Residuals of JBE frequency derivatives with respect to a
constant braking index law fit with $n = 2.5$, calculated over the
whole JBE interval (from 1988 May through 2015 June 15th).
The vertical gray lines denote the occurrence of glitches as reported by \citet{b35} and
http://www.jb.man.ac.uk/$\sim$pulsar/crab.html (see Table~\ref{TAB3}). Nine arrows
delimit chosen episodes and are placed at times when
(major) glitches appear to change the braking index.
The two dashed vertical lines mark the beginning and the end of the
data set. The gray broken line indicates the second derivative of the
continuous phase function defined in the text. Points used for the
fit with the braking law model $\varphi_j(t)$ are displayed in red
and green, while points in black are excluded from the fit, as
explained in the text.
Some post glitch residuals with the value below $-0.5\times 10^{-12}s^{-2}$ go
beyond the scale and are not shown. } \label{Fig3}
\end{figure*}

That the phase history of the Crab pulsar can be approximated
by a BIM to within one turn during a 14-month period
raises the question of
 how long such a description can go on? This idea can best be analyzed by using the publicly available JBE Crab pulsar ephemerides since 1988 because they represent the most complete and uniform description of Crab rotational history, which has been shown to be consistent with data in other wavebands.
As a first step in finding periods during which the braking index may be sufficiently constant, we plot
residuals of JBE-published frequency derivatives with respect to
frequency derivatives predicted by a BIM with $s=\frac{n-2}{n-1}=1/3$ ($n=2.5$)  and the parameters $a$ and $b$ in equation (\ref{EQN1})
chosen by a least squares fit minimizing residuals ${\dot f}_{res} = {\dot f}_{JBE}-\dot f$.
These residuals are shown in Fig.~\ref{Fig3}.

The graph clearly shows that on average,
the $n = 2.5$ braking index law describes the 26-year phase
history reasonably well. However, large systematic changes in
slope immediately after some (major) glitches signal that during
such periods, the braking index must be different from the average
value. Therefore, we divide the 26-year interval into ten subintervals, henceforth called episodes (see Fig.~\ref{Fig3}), each corresponding
to periods characterized by the absence of pronounced
changes in the braking index.
We expect that the phase history during an episode can be largely approximated by a constant braking index law,
as in the case of the interval of comparison of optical and radio data. At this stage in our discussion, the exact dates of the ten episodes are not yet well defined.
In the next sections we show how they can be better constrained.

To be able to properly categorize a glitch as a discontinuous change in frequency and frequency derivative, one would need densely distributed high resolution optical data from such an event. Unfortunately, they are not available yet.
To get an idea of how the spin-down behaves before and after the glitch event, we used the public available data (\citet{b2}  and http://www.jb.man.ac.uk/$\sim$pulsar/crab.html) and built the phase history since May 15, 1988, assuming that the phase prediction by the suggested third-degree polynomial is accurate to a (small) fraction of a period during each ephemerides epoch and that the phase is continuous between successive periods. We built up a table of pulsar's integer radio phase at the exact Julian dates of pulse arrival times given by JBE.
This table {\bf T} is a table of entries of
main pulse arrival times and integer number of turns the pulsar
has made from the starting date.  This table is broken into ten episodes according to boundaries shown in Fig.~\ref{Fig3} and determined as explained below (see also Appendix~\ref{appendixA}).

Inside each episode we seek a braking index law
$\varphi_j(t)$ in the same form as in Eq.~\ref{EQN1} with $s_j$ as an
additional fitting parameter. Our goal is to obtain the $s_j$ for the braking law that fits the particular episode, in such a way that it gives the smallest residuals over the whole episode. It turns out that this goal can only be achieved by using data in the apparently quiet part of the episode,  disregarding scattered data
immediately after the glitch at the beginning of the episode. Data
points whose phase is not considered in the fit are colored black in
Fig.~\ref{Fig3} (we comment on this choice below). In this way we obtain local fits
$\varphi_{j}(t)=c_j+a_j(1+b_j t)^{s_j}$, valid on
complete intervals of episodes $\lbrace T^b_j,T^b_{j+1}\rbrace$, where $T^b$ is the starting MJD for each episode.

\begin{figure*}
\centering
\includegraphics[width=17cm]{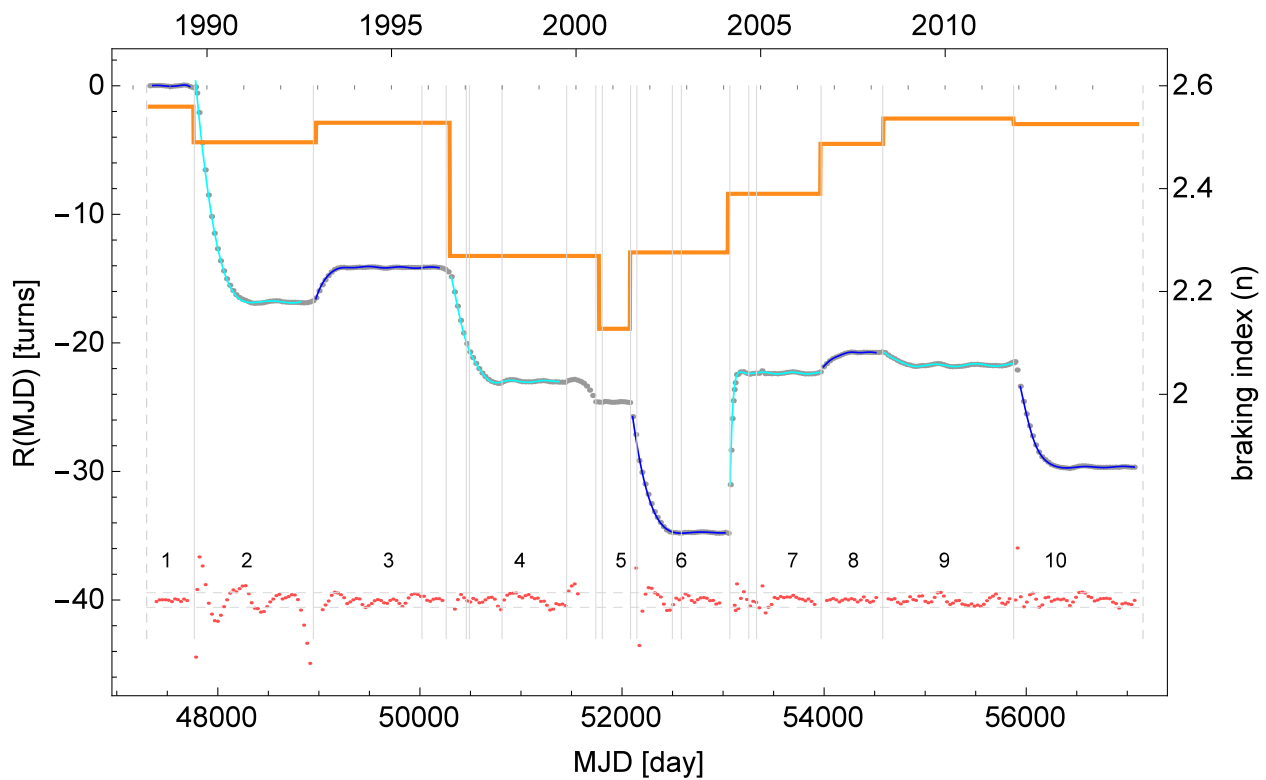}
\caption{JBE phase residuals $R(t)$ with respect to local fits
$\varphi_j(t)$ calculated for the 10 chosen episodes (see Fig.~\ref{Fig3}). Gray vertical lines are plotted at the dates of reported glitches \citep{b35}. The braking index
during episodes is shown in orange.
 {Sinusoidal fits to residuals, discussed below, are shown in blue and cyan to distinguish episodes.  Red} points at the bottom show ten times the difference between $R(t)$ and sinusoidal fits $R_j(t)$ at data points; the horizontal dashed lines bracket these final residuals with their standard deviation of 0.057 turns; the distribution of final residuals has wider wings than a Gaussian.}
\label{Fig4}
\end{figure*}

These local fits are joined into
a continuous curve $\Phi(t)$ by choosing $c_{j+1}$
in such a way that $\varphi_{j+1}(T^b_{j+1}) = \varphi_{j}(T^b_{j+1})$.
Residuals of the fit ($R(t) = {\bf T} - \Phi(t)$) are shown in  Fig.~\ref{Fig4} at all data points, and the parameters of $\varphi_j(t)$,
{ expressed as a Taylor series expansion of Eq.~(\ref{EQN1})}, are given in Table~\ref{TAB1}.
 When varying the parameters within the reported errors, the solution deviates from the best fits by less than 0.001 turn.\footnote{ As noted before, the split $\bf T = \Phi(t)+R(t)$ between $\Phi$ and $R$ is not unique. The one presented here is the result of our attempts to find one, where the contribution of $R(t)$ is as small and regular as we can find.}
The fits to those residuals, discussed in the next section, are also shown as blue and cyan curves on top of the R(t) dots in Fig.~\ref{Fig4}
(dots also include black data points in Fig.~\ref{Fig3}).
\begin{table*}
\begin{center}
\resizebox{\hsize}{!}%
{
\begin{tabular}{r c c c c c c c c c c c c}
\hline
j       &       $T^b_j\ [MJD]$ &   $\psi_j [10^9]$  &  $\delta\psi_j$ & $\nu_j\ [s^{-1}]$  &    $\delta \nu$ & $\dot \nu_j \ [10^{-10}s^{-2}]$ &$\delta \dot \nu$ & $\ddot \nu_j \ [10^{-20}s^{-3}]$ & $\delta \ddot \nu$ & $\dddot \nu_j \ [10^{-31}s^{-4}]$&$n_j$     & $\delta n$ \\
\hline
1       &       47327   &       0.080308783914  &       1.      &       29.98335828167  &       3.      &       -3.785953271    &       14.     &       1.223496        &       12.     &       -6.3    &       2.559365        &       24.\\
2       &       47759   &       1.199168006987  &       1.      &       29.96923749017  &       1.      &       -3.782862267    &       1.8     &       1.189028        &       0.52    &       -5.9    &       2.490158        &       1.1\\
3       &       48971   &       4.335378859396  &       1.      &       29.92968949582  &       0.9     &       -3.770288188    &       1.5     &       1.200702        &       0.40    &       -6.1    &       2.528066        &       0.85\\
4       &       50294   &       7.754097684269  &       1.      &       29.88667114924  &       0.8     &       -3.757375987    &       1.2     &       1.071861        &       0.29    &       -4.7    &       2.269065        &       0.61\\
5       &       51771   &       11.564963729546 &       1.      &       29.83880913666  &       4.      &       -3.744175564    &       28.     &       0.9995258       &       32.     &       -4.1    &       2.127468        &       67.\\
6       &       52080   &       12.361454906619 &       1.      &       29.82881752567  &       1.      &       -3.742269975    &       2.9     &       1.068559        &       1.0     &       -4.7    &       2.275959        &       2.2\\
7       &       53049   &       14.857460749680 &       1.      &       29.79752549720  &       1.      &       -3.735298384    &       3.2     &       1.119064        &       1.2     &       -5.3    &       2.389927        &       2.6\\
8       &       53962   &       17.206823673457 &       1.      &       29.76809511938  &       2.      &       -3.726586977    &       6.9     &       1.160253        &       3.9     &       -5.7    &       2.487030        &       8.3\\
9       &       54584   &       18.806047084704 &       1.      &       29.74808502226  &       0.9     &       -3.720427194    &       1.6     &       1.180117        &       0.43    &       -6.0    &       2.536288        &       0.93\\
10      &       55874   &       22.119341521240 &       1.      &       29.70669287202  &       0.9     &       -3.708385813    &       1.8     &       1.169154        &       0.51    &       -5.9    &       2.525551        &       1.1\\
\hline
\end{tabular}%
}
\end{center}
\caption{{ Parameters of local fits for different j episodes, expressed in the form of a Taylor series of the form $\varphi_j(t)=\psi_j+\nu_j t+\frac{1}{2}\dot \nu_j t^2+\frac{1}{6}\ddot \nu_j t^3 +\frac{1}{24}\dddot \nu_j t^4$ where the coefficients are constrained to fit the braking index law: i.e., this expression is a Taylor series expansion of Eq.~(\ref{EQN1}).
Quoted errors refer to the last reported digit.
Time in the argument of $\varphi_j$ starts at 00 UT of the appropriate $T^b_j$. The last two columns give the corresponding values of the braking index and its error.}}
\label{TAB1}
\end{table*}
Residuals in Fig.~\ref{Fig4} show that, during each episode, it is possible to find a local braking law
with respect to which $R\left (t\right )$ is never more than a few turns, even if the { pulsar makes billions of turns during the same period.}
It must be emphasized that residuals that are as small as those shown in Fig.~\ref{Fig4} can only be obtained by fitting
the phase on different episodes {\bf T}$_{j}$ with markedly different braking law indices, varying from
$\sim 2.1$ to $\sim 2.6$.

As already mentioned in the discussion of fitting optical data (see Fig.~\ref{Fig1}), the precise value of the braking
index on the episode may depend on the selection of data points after the glitch. Looking at Fig.~\ref{Fig3},
different choices for the black, green, and red points can possibly be made.
However, the requirement that phase residuals be as small as only a few turns very clearly narrows down an already narrow range of acceptable values of the braking index for each episode.
Finally, since $\Phi(t)$ is a continuous function, the residuals $R(t)$ must also be continuous at the boundaries between
episodes. This requirement narrows down the dates of boundaries between episodes to values $T^b_j$ as listed in Table~\ref{TAB1}.
\section{The rotational phase history of the Crab pulsar}

It is remarkable that the evolution of the rotational phase of the Crab
pulsar can be split into two parts: a regular phase $\Phi(t)$
consisting of a constant braking index during a given episode, but different
for each episode, and a small residual phase $R(t)$ that, during
all this time, wanders by no more than 35 turns, to be compared with
the $\sim 2.5\times 10^{10}$ turns the pulsar has made during the
9800 days covered by the JBE (see Fig.~\ref{Fig4}).

\begin{figure*}
\centering
\includegraphics[width=17cm]{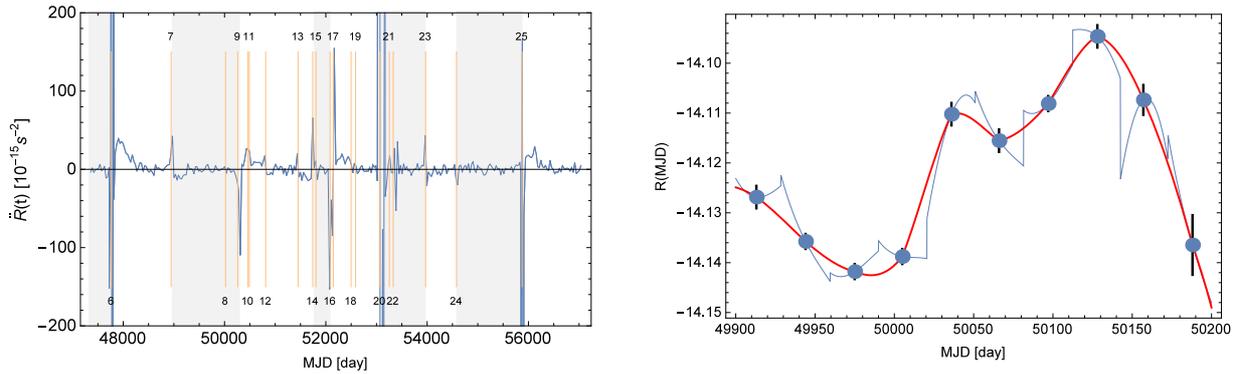}
\caption{Left: Second time derivative of the residual phase $R_{s}(t)$, a third-order spline fit through phase residuals from Table $\bf{T}$. The orange vertical lines are drawn at dates of reported glitches and labeled according to Table~\ref{TAB3}. Episodes are shaded intermittently as light gray and white.
Right: Phase residuals as a function of time for a short interval during episode 3. The red line shows $R_s(t)$, while the blue line is calculated as suggested by explanatory notes of JBE (see text for details). Error bars show JBE quoted arrival-time uncertainties.
}
\label{Fig5}
\end{figure*}

According to the dates reported in JBE  and in \citet{b34}, jumps in braking index are clearly related to large glitches
(see Table A1), while small glitches do not leave a notable imprint on the curve  $R(t)$. To recover the significance of small glitches from JBE, we calculated the second derivative of the function $R_{s}(t)$, which is constructed as a differentiable function from tabular values of $R(t)$ by cubic spline interpolation. The result is shown in Fig.~\ref{Fig5} (left panel). In this representation, small glitches, except possibly a few (n.\ 10, 11, 13), show up as spikes barely above the general noise in ${\ddot R}_{s}(t)$.
Only nine glitches (n. 6, 7,  9,  14, 16, 20, 23, 24, 25) show second-derivative ${\ddot R}_s(t)$ beyond  $2\times 10^{-14}s^{-2}$, which is smaller than the second derivative of optical phase residuals derived from the January 2009 data (${\ddot R}_{Jan}\approx 10^{-13}s^{-2}$ see Fig.~\ref{Fig2}). All of them are clearly related to a change in the braking index (Fig.~\ref{Fig4}).
{ The righthand panel of Fig.~\ref{Fig5} shows phase residuals $R_{s}(t),$ together with phase residuals
calculated as suggested by explanatory notes of JBE. Breaks (a few thousands of a turn) between the ephemerides dates, occur because the suggested expression for calculating the phase increment,
\begin{equation}
\varphi(t)=\varphi(t_0)+\frac{1}{P}(t-t_0)-\frac{1}{2}\frac{\dot P}{P^2}(t-t_0)^2+\frac{1}{3}\frac{{\dot P}^2}{P^3}(t-t_0)^3
\label{Eq3}
,\end{equation}
(P is the nominal period) does not yield an exact integer number of turns between two entries in Table $\bf T$. We suspect that the actual pulsar phase noise variation \citep{b55} is
the dominant cause for those phase residuals breaks.
The truncated Taylor series in Eq.~(\ref{Eq3}) is probably not sufficient to take complete account of phase noise variation.
Intrinsic timing noise
has also been confirmed by optical observations (Fig.~\ref{Fig2}), and
according to \citet{b16}, the difference in frequency
derivative obtained from radio ephemerides and from optical observations
is consistent with the difference between the red and
blue curve in the righthand panel of Fig.~\ref{Fig5}.}

 The curves of residuals in Fig.~\ref{Fig4} all have a characteristic shape. It is customary to express functions describing sets of data with a linear combination of a certain Hilbert space basis (eigenvectors) and natural to choose such a basis so that the concrete experimental result can be described by the fewest components.
Having tried different possibilities (including the widely used combination of sinusoids $+$ decaying exponentials), we find that the residuals can be described by only two complex Fourier components:

\begin{equation}
\begin{aligned}
R_j(t)=\Psi_j+A_j e^{-\lambda_j (t-T^b_j)}\sin\left (\omega_j (t-T^b_j )+\delta_j\right) \\
  + {\overline A}_j e^{-{\overline \lambda}_j (t-T^b_j )}\sin\left ({\overline \omega }_j(t-T^b_j)+\overline {\delta}_j\right).
\label{EQN4}
\end{aligned}
\end{equation}

\begin{figure*}
\centering
\includegraphics[width=17cm]{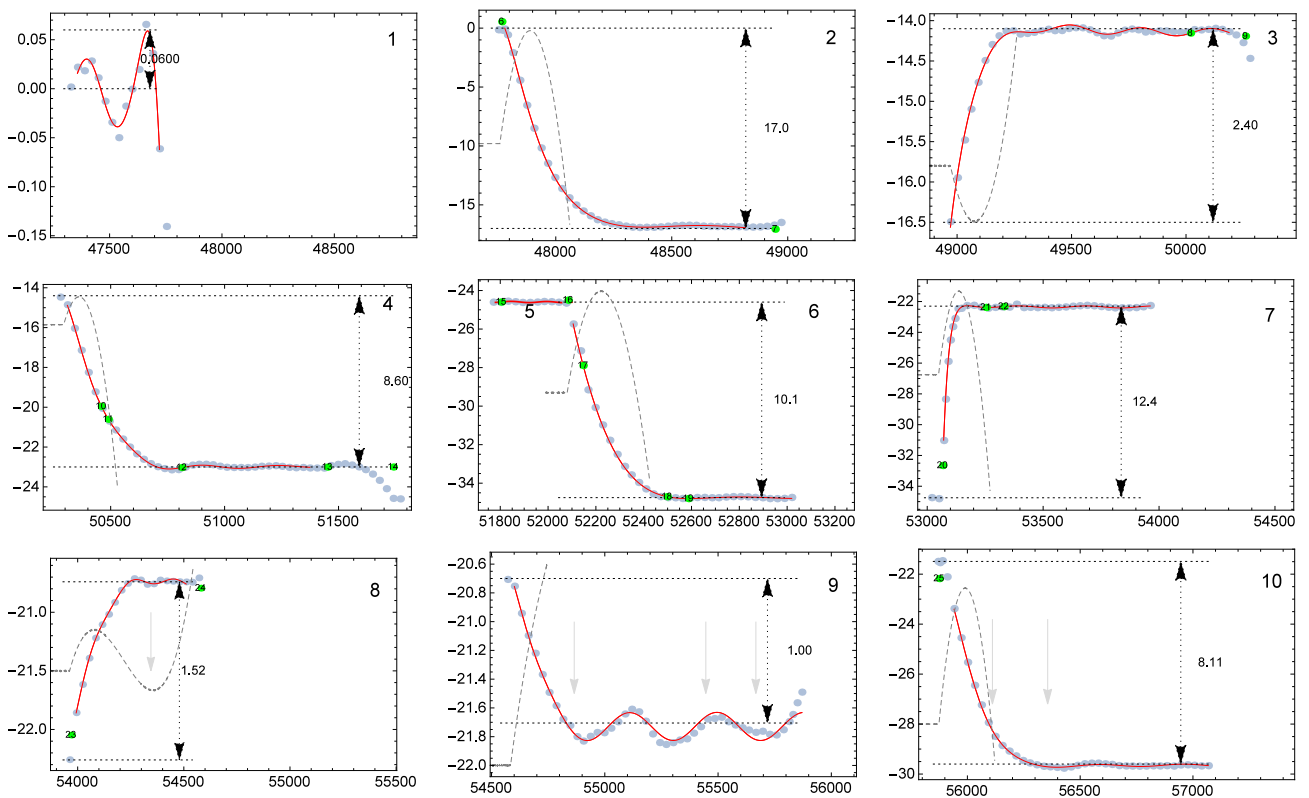}
\caption{Sinusoidal phase functions $R_j(t)$ (red lines) that fit
the JBE phase residuals (blue points) for episodes from $j$ = 1 to 10
(the short episode 5 is included in the graph showing episode 6). Time
on the abscissa is in MJD, and the scale is the same for all
episodes. Green dots are numbered as in \citet{b35} at dates of published glitches with the ordinate as calculated from $R_j(t)$. Ordinate scales are different and adjusted to
different amplitudes of oscillations. Each panel shows data of the complete episode and includes the first point of the next episode.
The difference between horizontal dotted lines represents a measure of the strength of the perturbation $R_j(t)$ caused by the glitch.
Dashed gray curves are the plots of  $\varphi_j(t)-\varphi_{j-1}(t)$, to show the difference of phases between two contiguous episodes. Gray arrows point to dates of the $> 100$ MeV X-ray
flares detected by Fermi \citep{b25,b27,b32} and AGILE
\citep{b26,b28}.} \label{Fig6}
\end{figure*}

We believe that this mathematical approach leads to a simpler description of pulsar noise.
The coefficients are given in Table~\ref{TBL2} and are sufficient to
produce a fit of $R(t)$ without systematic
trends in the residuals, as shown in gray at the bottom of Fig.~\ref{Fig4}. The standard deviation of the distribution of phase residuals is 0.057. It can be compared with the standard deviation of the set of fractional phase residuals between ephemerides epochs, which is 0.063. The declared arrival-time uncertainty in JBE is 0.01 or 300$\mu$s.  The details of these fits, together with some other pertinent information, are shown in Fig.~\ref{Fig6}.
\begin{table*}
\begin{center}
\resizebox{0.8\hsize}{!}%
{
\begin{tabular}{crrrrrrrrr}
\hline
j & $\Psi$ & $\omega \left [{\rm day}^{-1}\right ]$ & $\lambda \left [{\rm day}^{-1}\right ]$ &
$A$ & $\delta$ & $\overline \omega \left [{\rm day}^{-1}\right ]$ & $\overline \lambda \left [{\rm day}^{-1}\right ]$ &
$\overline A$ & $\overline \delta$ \\
\hline
 ~1 & -0.004 & 0.0345 & -0.0209 &   0.0000254 & -3.47  & 0.0236 & 0      & 0.034 & -0.0955\\
 ~2 & -16.91 & 0.0010 &  0.0113 &  177.00     & 0.100 & 0.0091 & 0      & 0.153 & -0.0392\\
 ~3 & -14.15 & 0.0010 &  0.0072 &   8.63      & -0.282 & 0.0206 & 0      & 0.047 & -2.9822\\
 ~4 & -22.98 & 0.0054 &  0.0092 &  14.4       & 0.621 & 0.0187 & 0.0018 & 0.320 &  2.7275\\
 ~6 &  36.72 & 0.0037 &  0.0018 &  12.7       & -2.46  & 0.0008 & 0      & 73.50 & -2.3370\\
 ~7 & -22.35 & 0.0010 &  0.0321 & 217.        & -0.103 & 0.0213 & 0      & 0.066 &  0.5691\\
 ~8 & -20.78 & 0.0068 &  0.0075 &   1.36      & -1.86  & 0.0306 & 0      & 0.029 & -1.0713\\
 ~9 & -21.72 & 0.0072 &  0.0131 &   1.68      & 0.73  & 0.0163 & 0      & 0.098 & -0.709\\
 10 & -29.65 & 0.0059 &  0.0102 &  16.1       & 0.48  & 0.0166 & 0      & 0.046 &  2.45\\
\hline
\end{tabular}%
}
\end{center}
\caption{Coefficients of the fitting function $R_j(t)$ in Eq.~\ref{EQN2}}\label{TBL2}
\end{table*}
The ansatz in Eq.~\ref{EQN4} is purely mathematical and is  an almost satisfactory description of phase residuals $R(t)$. A posteriori, we note that the first sinusoid has very low values of $\omega$ so is essentially a decaying component, while in all episodes but one, the second sinusoid is not damped.

{It should be understood that glitches may not be modeled exactly by our analysis because the time resolution of JBE data is not sufficient to describe the exponential decay of a glitch that lasts a few days \citep{b55}. However, because of all that has been said, the JBE data are reliable as to the global phase behaviour and on timescales longer than about a month. In this respect, the decaying component at the beginning of several  episodes resembles the usual after-glitch recovery behavior frequently modeled in the literature through a single-parameter exponential (e.g., \citealt{b51}, with a characteristic timescale of 320 days)}.

Figures~\ref{Fig4} and~\ref{Fig6}
clearly illustrate the meaning of the split of the phase behavior into the braking index part $\Phi(t)$ and residuals $R(t)$. It is quite clear that toward the end of an episode, residuals relax to the perfect braking law solution and eventually oscillate for years by only a very small fraction of a turn. On the other hand, as hinted in Fig.~\ref{Fig6}, the braking-law part of an episode  ($\varphi_j(t)$) generally differs quite considerably from the braking law of the previous episode ($\varphi_{j-1}(t)$).
In fact, if the difference $\varphi_j(t)-\varphi_{j-1}(t)$ was plotted to the end of episodes, it would reach values several tens or a hundred times the value $\Delta  N$. An exception is the short episode 8, which follows a relatively weak glitch. For this episode R(t) and $\varphi_j(t)-\varphi_{j-1}(t)$ are comparable, and therefore, the split between the braking index part and residuals is not as clear cut. Figure~\ref{Fig6} also confirms that, within limits allowed by the time resolution of available data, the breaks between episodes occur on the dates of reported glitches. It appears plausible to classify glitches into two groups: those for which the change in the phase $\Phi(t)$ dominates residuals $R(t)$ by many factors of ten (6, 7, 9, 16, 20, 23, 24, 25) and those where the integral change in $\varphi_j(t)-\varphi_{j-1}(t)$ is comparable or insignificant with respect to the amplitude of $R(t)$ (all other glitches beyond \#6).

\section{Discussion and conclusions}

Our analysis shows that the phase evolution of the Crab pulsar can be described as
a series of constant braking-law episodes, with the
braking index changing abruptly after each episode in the range of
values between 2.1 and 2.6. Phase residuals with respect to such a
smooth phase description amount to only a few turns in $\sim 10^9$ turns executed during an episode. The split between the smooth braking-law-dominated part and residuals is not
mathematically unique, but requirements that phase residuals be
as small as only a few turns and that the phase between ephemerides epochs clearly converges to the characteristic braking index solution of the episode narrow the choice of braking index parameters.

A similar conclusion concerning the behavior of the braking index has been recently obtained by \cite{b51} from an independent analysis of 45 years of radio data on the rotational history of the Crab pulsar. Results obtained from a fit with a single Taylor series returns a behavior of $f$ and $\dot f$ (their Figures 1 and 2) very similar to the one shown in Figure~\ref{Fig3} (with a best fitting $n=2.34$). Variations in the braking index in the period between 1996 and 2006, characterized by a high concentration of glitches, were also noted by \cite{b51}.
While they consider it a weak, unexplained effect on the background of the previous rotational history
(described by a simple slowdown with braking index 2.519(2)), we offer here a different
interpretation that can account for the overall timing irregularities of the Crab pulsar.

\begin{figure*}
\centering
\includegraphics[width=17cm]{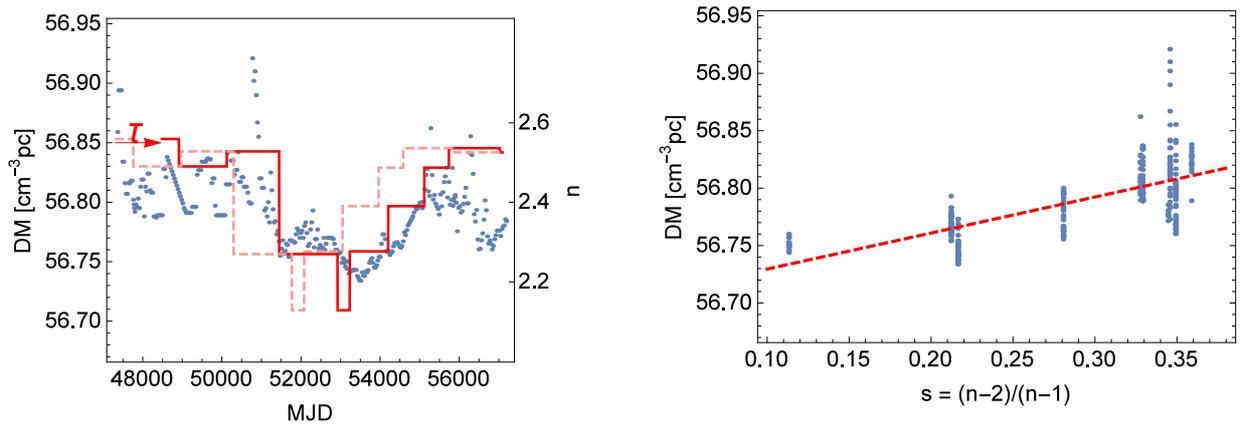}
\caption{a)~Left: dispersion measure $DM(t)$ (blue dots), braking
index $n(t-\tau)$ for $\tau = 1100$ days (continuous red line). The dashed red line shows $n(t)$.
 b)~Right: correlation between braking index
($n$) and dispersion measure $DM$. The correlation coefficient is
0.7.} \label{Fig7}
\end{figure*}

According to our interpretation, glitches
and abrupt changes in the braking mechanism may be part of the same
physical process that also drives semi-periodic timing noise
between glitches (Fig.~\ref{Fig4}).
In fact, JBE data provide an interesting correlation of the
braking index and dispersion measure. Namely, the dispersion
measure (as listed in JBE) follows the braking index with a time
delay of 1100$^{+450}_{-250}$ days as shown in Fig.~\ref{Fig7}.
The delayed response of dispersion measure to the value of braking
index lends support to the idea of \citet{b11} that change of the
braking index has to do with a pulsar-wind-driving torque and is
also consistent with the idea that eddy currents threading the
pulsar nebula ionize it and thus inject varying amounts of free
electrons.

From this new perspective, glitches, timing noise, changes in braking
torque, and dispersion measure appear to be part of a common
mechanism that jumps between different braking modes. Distinctly
long timescales of timing noise oscillations suggest that the
mechanism can hardly be connected to phenomena occurring within the
pulsar. It appears plausible that the ``instantaneous'' change in
braking torque is caused by some instability, which varies the
configuration of the external electromagnetic field and currents in
the nebular plasma through which the pulsar interacts with its
nebula. Such an interaction is needed in order to understand the
acceleration and energizing mechanisms of the pulsar nebula
\citep{b19,b20}, such as the highly dynamical flow of relativistic
particles in the form of equatorial wind and polar jets, as seen in
Hubble Space Telescope and Chandra images \citep{b21}.

The possible
occurrence of plasma instabilities causing ``instantaneous'' changes
in the braking mechanism is expected to produce observable
changes in radiation from the Crab nebula. It is tempting to
consider the possibility \citep{b24} that plasma instabilities
occurring through magnetic field line reconnection drive the
recently observed gamma ray flares \citep{b25,b26,b27,b28,b32}.
Arrows in Fig.~\ref{Fig6} point to instants when the six observed
flares occurred in the Crab nebula. The same mechanism, which also
appears to be acting in the solar corona to produce gamma ray flares
\citep{b000}, may be able to provide a sudden short release of
braking torque by disconnecting the magnetic field lines from the
pulsar from those threading the nebular plasma. The same mechanism
may also be responsible for sharp increases in dispersion measure
(Fig.~\ref{Fig7}, left) by emitting highly energetic particles into
the neutral nebula, thus ionizing it.

The back reaction of plasma instabilities on the pulsar, possibly
associated to these flares, has not been studied yet, but it does not seem to be simultaneous with the occurrence of the flare
\citep{b29,b30}. In view of the delayed correlation between
dispersion measure and braking index, this appears understandable --
perturbations caused by magnetic reconnection travel long distances
before reaching the pulsar or before permeating the nebula. Nevertheless, the
interflare time inferred from numerical simulations (hundreds of
days) \citep{b22,b23,b24} appears to be broadly consistent with the
timescales observed in oscillations of JBE phase residuals. It is
quite remarkable that the occurrence of the reported gamma-ray
flares is also consistent with such a timescale.

Our ultra-fast optical observations of the Crab pulsar with Aqueye
and Iqueye stimulated the line of research presented here.
Future simultaneous radio and optical timing measurements, as well as
optical imaging and gamma-ray observations, will be crucial for
revealing the source of the braking mechanism, in particular if it is
located in the external electromagnetic field through which the
pulsar interacts with the surrounding plasma, as suggested by our
results.

\begin{acknowledgements}

This work is based on observations made with ESO Telescopes at the
La Silla Paranal Observatory under program IDs 082.D-0382 and
084.D-0328(A) and on observations collected at the Copernico
Telescope (Asiago, Italy) of the INAF-Osservatorio Astronomico di
Padova. We acknowledge the use of the Crab pulsar radio ephemerides
available on the web site of the Jodrell Bank Radio Observatory
(http://www.jb.man.ac.uk/$\sim$pulsar/crab.html, see \citealt{b2}). This
research has been partly supported by the University of Padova under
the Quantum Future Strategic Project, by the Italian Ministry of
University MIUR through the program PRIN 2006, by the Project
of Excellence 2006 Fondazione CARIPARO, and by INAF-Astronomical Observatory of Padova.
One of us (A. \v C) would like to express his gratitude to the relativity group at the Silesian University in Opava for their support and friendship.
The authors wish to thank the referee, Patrick Weltevrede, for his valuable
comments that helped to improve the paper.
\end{acknowledgements}

\begin{appendix}
\section{Crab phase data}
\label{appendixA}

We construct the decadal timing solution of the
Crab pulsar as a series of pairs $\bf{T}_{j} = \{\rm{MJD}_{j} +
(1/86400) \rm{TOA}_{j}, z_{j}\}$, where $\rm{MJD}_{j}$ is the mean
Julian date of the $j$-th entry of JB ephemerides, $\rm{TOA}_{j}$ is
the first arrival time of the pulse at $\rm{MJD}_{j}$ (the
t$_{\rm{JPL}}$ entry in JBE), and $z_{j}$ is the integer number of
turns the pulsar has made from the starting date May 15, 1988
(MJD$_{1}$=47296). Integers $z_{j}$ are calculated by summing the
integer number of turns N$_{j}$ that the pulsar has made between
MJD$_{j-1}$ and MJD$_{j}$ ($z_{j} = \sum{}^{j}_{k=1}N_{k}$). N$_{k}$
is calculated using the JB published values of frequency and
frequency derivative. Using the formula suggested by explanatory notes to JBE, we calculate it as follows.

\begin{figure*}
\centering
\includegraphics[width=17cm]{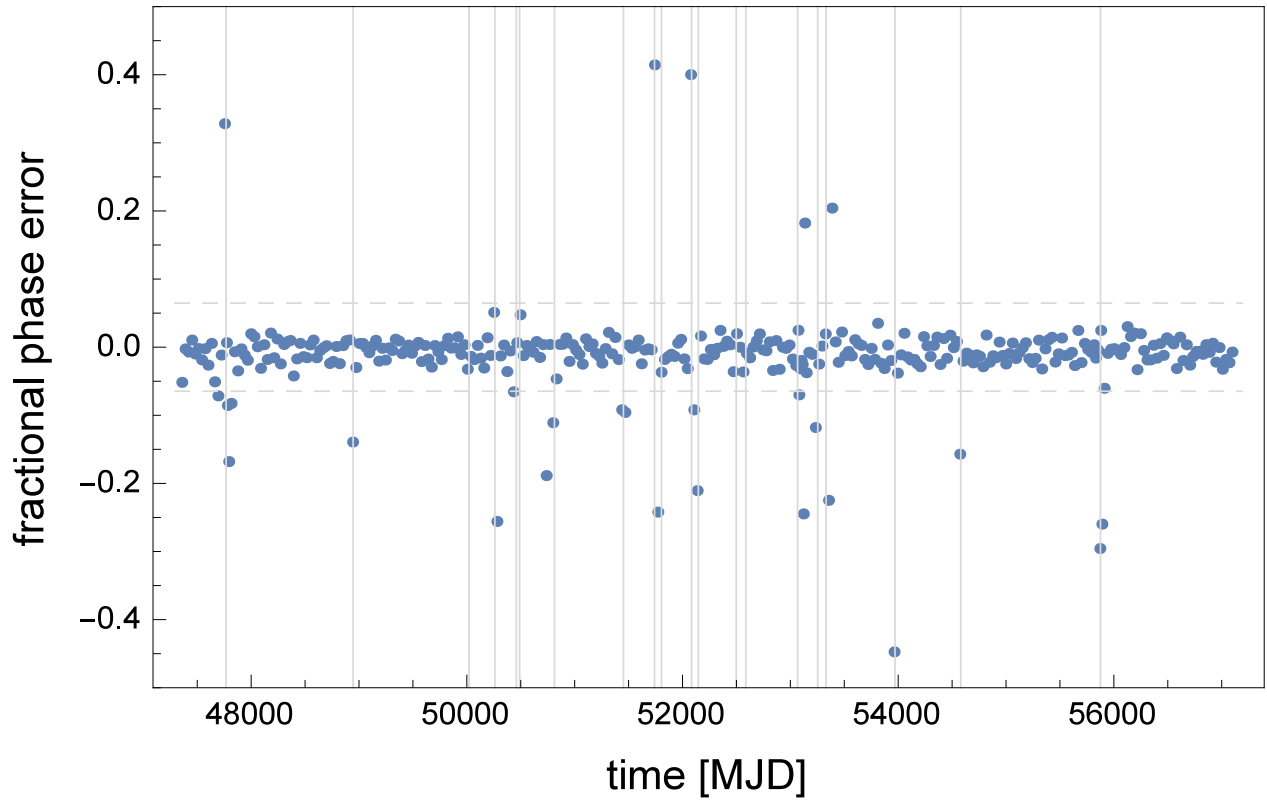}
\caption{Fractional part of calculated number of turns between successive TOAs' fractional phase error. Vertical lines denote the dates of reported glitches, starting from glitch no.\ 5 in Table~\ref{TAB3}. { Horizontal dashed lines are at $\pm 0.062$, the variance of the (wide winged) phase error distribution.}}
\label{Fig8}
\end{figure*}

 Let $\Delta t =43200\left ({\bf T}_j-{\bf T}_{j-1}\right )$, and let $\nu_j$ and ${\dot \nu}_j$ be the frequency and frequency derivative listed at the j-th ephemerides entry in JBE. Define $P_j=1/\nu_j$ and ${\dot P}_j=-{\dot \nu_j}{\nu_j}^{-2}$, then $\Delta N=\left(\frac{1}{P_{i-1}}+\frac{1}{P_i}\right)\Delta t-\left(\frac{\dot P_{j-1}}{{P_{j-1}}^2}-\frac{\dot P_{j}}{{P_{j}}^2}\right )\frac{\Delta t^2}{2}+\left(\frac{{\dot P_{j-1}}^2}{{P_{j-1}}^3}+\frac{{\dot P_{j}}^2}{{P_{j}}^3}\right )\frac{\Delta t^3}{3}$ and $N_j=$IntegerPart$[\Delta N]$. The evaluation of $\Delta N$ in most cases yields an integer, as it should. However, in 25
cases, almost all of them occurring at the time of a glitch, the
calculated number of turns between successive TOA's has a fractional
part that is inconsistent with the quoted TOA accuracy. The fractional phase errors are shown in Fig.~\ref{Fig8}.
\begin{table*}
\begin{center}
{
\begin{tabular}{|l |l |c |c |c|}
\hline
no &    glitch  & episode & $T^b$  & $\delta f/f$ \\
   &   [MJD] & no. & [MJD] & [$10^{-9}$]  \\
\hline
~1&    40491.8    & &    &       7.2\\
~2&    41161.98& &  &           1.9\\
~3&    41250.32& &  &           2.1\\
~4&    42447.26& &  &      35.7\\
~5&    46663.69& &  &           6.0\\
  &            & 1 &47327  &\\
~6&    47767.504&  2 &  47759&    81\\
~7&    48945.6    & 3 &    48971&   4.2\\
~8&    50020.04& &   &     2.1\\
~9&    50260.031&   4 &50294&    31.9\\
10&    50458.94& & &       6.1\\
11&    50489.7    & &  &          0.8\\
12&    50812.59& & &       6.2\\
13&    51452.02& & &       6.8\\
14&    51740.656& 5 &  51771&    25.1\\
15&    51804.75& & &        3.5\\
16&    52084.072& 6  & 52080&    22.6\\
17&    52146.758& &  &      8.9\\
18&    52498.257& &  &      3.4\\
19&    52587.2    & &   &         1.7\\
20&    53067.078&  7 & 53049&    214\\
21&    53254.109& & &        4.9\\
22&    53331.17& & &       2.8\\
23&    53970.19&   8 & 53962&    21.8\\
24&    54580.38&  9 &  54584&     4.7\\
25&    55875.5 &  10&  55874& 49.2\\
\hline
\end{tabular}
}
\end{center}
\caption{Numbers and dates of glitches (from \citet{b35}). The MJD episode is the starting date of the episode as listed in Table~\ref{TAB1}, and the last column quotes reported frequency jumps during the glitch.}
\label{TAB3}
\end{table*}

Table A.1 lists all glitches reported in \citet{b35}
(http://www.jb.man.ac.uk/pulsar/glitches.html) and
the starting dates of episodes 2 to 10.

\end{appendix}
\end{document}